\documentclass[reqno,11pt]{amsart}
\usepackage{amsmath,amssymb,bbm,color}

\def\build#1_#2^#3{\mathrel{\mathop{\kern 0pt#1}\limits_{#2}^{#3}}}
\newcommand{\madebyTeXmacs}{\footnote{This text has been produced using GNU T\kern-.1667em\lower.5ex\hbox{E}\kern-.125emX\kern-.1em\lower.5ex\hbox{\textsc{m\kern-.05ema\kern-.125emc\kern-.05ems}} (see {\tt http://www.texmacs.org}).}}
\newcommand{\withTeXmacstext}{This text has been produced using GNU T\kern-.1667em\lower.5ex\hbox{E}\kern-.125emX\kern-.1em\lower.5ex\hbox{\textsc{m\kern-.05ema\kern-.125emc\kern-.05ems}} (see {\tt http://www.texmacs.org})}
\newcommand{\tmop}[1]{\ensuremath{\operatorname{#1}}}

\begin{document}

\title{On the Surface Tensions of Binary
Mixtures
} 
\author{
Jean RUIZ 
}

\begin{abstract}
  For binary mixtures with fixed concentrations of the species, various
  relationships between the surface tensions and the concentrations are
  briefly reviewed. 
  
 \vspace{.5cm} 
  
{\noindent}  
\textit{Preprint CPT-2004/P.059}
\end{abstract}

 \maketitle

\renewcommand{\thefootnote}{}
\madebyTeXmacs
\renewcommand{\thefootnote}{\arabic{footnote}}

\renewcommand{\thefootnote}{}
\footnote{Talk presented at the $8^{th}$ ICCMSP, September $21-24$ $2004$ Marrakech.}
\renewcommand{\thefootnote}{\arabic{footnote}}

\renewcommand{\thefootnote}{}
\footnote{ CPT-CNRS, Luminy Case 907
  F-13288, Marseille Cedex 9, France.
  \\
 \indent E-mail adress: Jean.Ruiz@cpt.univ-mrs.fr}
\renewcommand{\thefootnote}{\arabic{footnote}}

\setcounter{footnote}{0} 

\thispagestyle{empty}

When we consider a binary mixture of two chemical species 1 and 2, in
equilibrium with its vapor, one of the problems, experimentally as well as
theoretically, is to predict how the corresponding surface tension depends on
the composition of the mixture. Some relationship is expected which would give
this surface tension, as an interpolation between the surface tensions of  the
two species when they are chemically pure. In this note, we briefly discuss
various relation-ships based on thermodynamical  considerations as well as
other ones obtained more recently in the frame of Solid-On-Solid models of
interfaces {\cite{DR}} and bulk statistical mechanical models of binary
mixtures {\cite{DMR}}.

\vspace{.2cm}
\noindent
\textbf{Thermodynamical se{\noindent}mi-empirical equations}

Let us use $\text{$\tau_{( 1, 2 ) |0}$}$ to denote the surface tension of a
mixture of two species $1$ and $2$ and let $\text{$\tau_{1|0}$ and
$\tau_{2|0}$}$ be the surface tensions of each species. In this section, we
present several equations which have been derived, according to different
assumptions, by using thermodynamical considerations.

For ideal or nearly ideal solutions, a fairly simple treatment, due to
Guggenheim {\cite{Gu}}, leads to the following equation
\begin{equation}
  \label{eq:gug} e^{- \beta a \tau_{( 1, 2 ) |0}} = c_1 e^{- \beta a
  \tau_{1|0}} + c_2 e^{- \beta a \tau_{2|0}}
\end{equation}
where $c_1$ is the fixed molar fraction of species $1$ in the $( 1, 2 )$
mixture, $c_2 = 1 - c_1$, the fixed molar fraction of species $2$, $a$ is the
mean surface area per molecule, and $\beta = 1 / kT$ is the inverse
temperature.

A very simple relationship for the so called regular solutions comes from
Prigogine and Defay {\cite{DP}}, who proposed the equation
\begin{equation}
  \label{depr} \tau_{( 1, 2 ) |0} = c_1 \tau_{1|0} + c_2 \tau_{2|0} - Kc_1 c_2
\end{equation}
with $K$ a semiempirical constant.

A simple treatment due to Eberhart {\cite{Eb}} assumes that the surface
tension of a binary solution is linear in the surface composition, that is
\begin{equation}
  \label{eq:eb} \tau_{( 1, 2 ) |0} = c^s_1 \tau_{1|0} + c^s_2 \tau_{2|0}
\end{equation}
where the $c^s_i$, $i = 1, 2$, denote the mole fraction near the surface of
phase separation, and that the ratio $c^s_1 / c_1$ is proportional to the
ratio $c_2^s / c_2$.

Finally, when the surface tensions $\tau_{1|0}$ and $\tau_{2|0}$ differ
appreciably, a semiempirical equation attributed to Szyszkowsky ({\cite{Sz}},
{\cite{Sz2}}) gives:
\begin{equation}
  \label{eq:sz} \frac{\tau_{( 1, 2 ) |0}}{\tau_{1|0}} = 1 - B \ln \left( 1 +
  \frac{c_2}{A} \right)
\end{equation}
where two characteristic constants $A$ and $B$ of the compounds have been
used, and $c_2$ is the concentration of the species with the smaller surface
tension.

We refer the reader to Adamson's book {\cite{Ad}} (Chapter III, Section 4),
and references therein, for a detailed discussion of the above equations.

Let us  also mention that an extensive development for various types of non
ideal solutions  has been made by Defay, Prigogine and co--workers: see the
monography {\cite{DPBE}}.

\vspace{.2cm}
\noindent
\textbf{Solid-On-Solid (SO{\noindent}S) models}

Let us first consider the SOS model in dimension $d = 2$. We let $h_0, \ldots,
h_N \in \mathbb{R}$ be a collection of heights describing an interface. For
simplicity, we assume that the energetic cost of the interface is proportional
to its length, namely given by
\begin{equation}
  H ( J_{\alpha} ; h_0 \cdots h_N ) = J_{\alpha} \sum^{N - 1}_{i = 0} [ 1 +
  |h_{i + 1} - h_i | ]
\end{equation}
where $J_{\alpha}$ represents the energetic cost per unit length for the
interface. The associated density of free-energy or interfacial tension is
then defined by
\begin{equation}
  \tau_{} ( \beta J_{\alpha} ) = - \frac{1}{\beta} \lim_{N \rightarrow \infty}
  \frac{1}{N} \ln Z ( N, \beta J_{\alpha} )
\end{equation}
where $Z ( N, \beta J_{\alpha} )$ is the following partition function
\[ Z ( N, \beta J_{\alpha} ) = \int^{+ \infty}_{- \infty} dh_0 \cdots \int^{+
   \infty}_{- \infty} dh_N e^{- \beta H ( J_{\alpha} ; h_0 \cdots h_N )}
   \delta ( h_0 - 0 ) \delta ( h_N - 0 ) \]
The quantities $\tau_{} ( \beta J_1 )$ and $\tau_{} ( \beta J_2 )$ will
represent the surface tensions of two species $1$, and $2$.

To model the interface for binary mixtures,  we will use the disordered
generalization of the SOS model. Namely, we consider that the coupling per
unit length is no more a constant but a random variable $J$. This variable may
take two values $J_1$ and $J_2$ with probabilities $c_1$ and $c_2 = 1 - c_1$,
where $c_1$ is physically the mole fraction of particles $1$ in the bulk of
the $( 1, 2 )$ mixture. This allows us to represent the interface by two
$\tmop{sets}$ of independent random variables $\{ h_0, \ldots, h_N \}$ and $\{
J^0, \ldots, J^{N - 1}_{} \}$. In this way, the interface can adjust its
height $h$ and, by moving or not molecules, also the corresponding energetic
cost $J$ taking into account the fixed concentration. In this approach the
molecules of the mixture are not distinguishable and have similar size. Each
site $i$ is occupied by one molecule. The energetic cost of this interface is
given by the Hamiltonian
\begin{equation}
  H ( J^0 \cdots J^{N - 1} ; h_0 \cdots h_N ) = \sum^{N - 1}_{i = 0} J^i [ 1 +
  |h_{i + 1} - h_i | ]
\end{equation}
where $J^i$ may take the value $J_1$ or $J_2$ with probability $c_1$ or $c_2 =
1 - c_1 .$ The associated partition function is given by
\[ Z_N ( J^0 \cdots J^{N - 1} ) = \int^{+ \infty}_{- \infty} dh_0 \cdots
   \int^{+ \infty}_{- \infty} dh_N e^{- \beta H ( J^0 \cdots J^{N - 1} ; h_0
   \cdots h_N )} \delta ( h_0 ) \delta ( h_N ) \]
It is well known that random systems are often related to disordered systems
for which one has introduced the notion of quenched and annealed disorder. For
the annealed case, the couplings are considered to be random and will be
treated in the same way than the heights. For the quenched disorder, the
couplings are frozen in a given configuration. There are then two ways to
define the associated free energy density
\begin{equation}
  \tau_{( 1, 2 ) |0}^{\text{quenched}} = - \frac{1}{\beta a} \lim_{N
  \rightarrow \infty} \frac{1}{N} \langle \tmop{lnZ}_N ( J^0 \cdots J^{N - 1}
  ) \rangle
\end{equation}
and
\begin{equation}
  \tau_{( 1, 2 ) |0}^{\text{annealed}} = - \frac{1}{\beta a} \lim_{N
  \rightarrow \infty} \frac{1}{N} \ln \langle Z_N ( J^0 \cdots J^{N - 1} )
  \rangle
\end{equation}
where the average $\langle \hspace{0.25em} \cdot \hspace{0.25em} \rangle$ has
to be taken with respect to the coupling distribution.

Using quenched disorder to compute the free energy, we obtain the following
equation to express the surface tension of the mixture like a convex
combination of the pure component surface tensions
\[ \tau_{( 1, 2 ) |0}^{\text{quenched}} = c_1 \tau_{} ( \beta J_1 ) + c_2
   \tau_{} ( \beta J_2 ) \]
This kind of formula is therefore valid when the system finds its equilibrium
position within the configurations for a given set of couplings.

The other approach, i.e. if we use the annealed disorder, leads to :
\[ e^{- \beta \tau_{( 1, 2 ) |0}^{\text{annealed}}} = c_1 e^{- \beta \tau_{} (
   \beta J_1 )} + c_2 e^{- \beta \tau_{} ( \beta J_2 )} \]
Here, the variables $J^i$ and $h_i$ are treated on an equal basis. This
implies that the molecules at the interface are sufficiently mobile to allow
the interface to adjust itself, both in heights and in composition.

\vspace{.2cm}
\noindent
\textbf{Bulk model of binary mixture}

This Section is devoted to a discussion of the problem within a lattice bulk
statistical mechanical model describing the binary mixture in equilibrium with
its vapor. Studies of various models of binary lattice gases can be found in
Refs. {\cite{WW,LG}}.

Here, we consider a lattice gas system with two kinds of particles, where each
lattice site can be in one of the three states, $0, 1, 2$, interpreted,
respectively, as an empty site, a site occupied by a particle of the first
kind of the model, and a site occupied by a particle of the second kind.
Whenever the particles $2$ are not allowed the system reduces to the usual
Ising model, in its lattice gas version, with coupling constant $J_1 / 2$, and
analogously, when particles $1$ are not allowed, it reduces to the Ising model
with coupling constant $J_2 / 2$. Namely, to each site $x \in \mathbb{Z}^d$,
$d = 2, 3$, we associate a variable $s_x$ taking values in the set $\Omega =
\left\{ 0, 1, 2 \right\}$. We will say that the site $x$ is empty when $s_x =
0$ and that it is occupied otherwise. The energy of a configuration
$\mathbf{s}_{\Lambda} = \left\{ s_x \right\}_{x \in \Lambda}$ in a finite box
$\Lambda \subset \mathbb{Z}^d$ is defined by
\begin{eqnarray}
  H_{\Lambda} ( \mathbf{s}_{\Lambda} ) & = & \sum_{\langle x, y \rangle
  \subset \Lambda} E ( s_x, s_y ) \label{eq:hamilt} \nonumber\\
  E ( s_x, s_y ) & = & J_1 \left[ \delta ( s_x, 1 ) \delta ( s_y, 0 ) + \delta
  ( s_x, 0 ) \delta ( s_y, 1 ) ] \right. \nonumber\\
  &  & + J_2 \left[ \delta ( s_x, 2 ) \delta ( s_y, 0 ) + \delta ( s_x, 0 )
  \delta ( s_y, 2 ) \right] \nonumber
\end{eqnarray}
where $\langle x, y \rangle$ denote nearest neighbour pairs, $\delta$ is the
usual Kronecker symbol $\delta ( s, s' ) = 1$ if $s = s'$ and $\delta ( s, s'
) = 0$ otherwise; $J_1$ and $J_2$ are positive constants. Notice that the bond
energy satisfies:
\begin{eqnarray*}
  &  & E ( 0, 0 ) = E ( 1, 1 ) = E ( 2, 2 ) = E ( 1, 2 ) = E ( 2, 1 ) = 0\\
  &  & E ( 0, 1 ) = E ( 1, 0 ) = J_1, E ( 0, 2 ) = E ( 2, 0 ) = J_2
\end{eqnarray*}

Fixed densities of the three species are introduced through the canonical
Gibbs ensemble of configurations $\mathbf{s}_{\Lambda}$ such that 
$$\sum_{x \in
\Lambda} \delta ( s_x, 0 ) = N_0, \ \sum_{x \in \Lambda} \delta ( s_x, 1 ) =
N_1, \ \sum_{x \in \Lambda} \delta ( s_x, 2 ) = N_2$$ 
where the sum  $N_0 + N_1
+ N_2$ equals  the number of sites of $\Lambda$. The associated partition
functions with boundary condition bc are given by
\[ Z_{bc} ( \Lambda ; N_1, N_2 ) = \sum_{\mathbf{s}_{\Lambda} \in
   \Omega^{\Lambda}} e^{- \beta H_{\Lambda} ( \mathbf{s}_{\Lambda} )} \delta
   \left( \sum_{x \in \Lambda} \delta ( s_x, 1 ), N_1 \right) \delta \left(
   \sum_{x \in \Lambda} \delta ( s_x, 2 ), N_2 \right) \chi^{bc} (
   \mathbf{s}_{\Lambda} ) \]
where $\chi^{bc} ( \mathbf{s}_{\Lambda} )$ is a characteristic function
standing for the boundary condition bc. We shall be interested in particular
to the following boundary conditions:
\begin{itemize}
  \item the empty boundary condition: $\chi^{\text{emp}} (
  \mathbf{s}_{\Lambda} ) = \prod_{x \in \partial \Lambda} \delta ( s_x, 0 )$
  
  \item the mixture boundary condition: $\chi^{\text{mixt}} (
  \mathbf{s}_{\Lambda} ) = \prod_{x \in \partial \Lambda} ( 1 - \delta ( s_x,
  0 ) )$
  
  \item the free boundary condition: $\chi^{\text{fr}} ( \mathbf{s}_{\Lambda}
  ) = 1$
\end{itemize}
where hereafter, the boundary $\partial \Lambda$ of the box $\Lambda$ is the
set of sites of $\Lambda$ that have a nearest neighbour in $\Lambda^c =
\mathbb{Z}^d \setminus \Lambda$.

The free energy per site corresponding to the above ensemble as a function of
the densities $\rho_1$ and $\rho_2$ of the particles $1$ and $2$ is
\begin{equation}
  \label{freeenergy} f ( \rho_1, \rho_2 ) = \lim_{\Lambda \uparrow
  \mathbbm{Z}^d} - \frac{1}{\beta | \Lambda |} \ln Z_{bc} ( \Lambda ; [ \rho_1
  | \Lambda | ], [ \rho_2 | \Lambda | ] )
\end{equation}
where $[ \hspace{0.25em} \cdot \hspace{0.25em} ]$ denotes the integer part and
the thermodynamic limit $\Lambda \uparrow \mathbb{Z}^d$ is taken in the van
Hoove sense {\cite{R1}}.

It is convenient to consider also a grand canonical Gibbs ensemble, which is
conjugate to the previous ensemble, and whose partition function, in the box
$\Lambda$ is given by
\begin{equation}
  \label{gcpf} \Xi_{bc} ( \Lambda ; \mu_1, \mu_2 ) =
  \sum_{\mathbf{s}_{\Lambda} \in \Omega^{\Lambda}} e^{- \beta H_{\Lambda} (
  \mathbf{s}_{\Lambda} ) + \mu_1 \sum_{x \in \Lambda} \delta ( s_x, 1 ) +
  \mu_2 \sum_{x \in \Lambda} \delta ( s_x, 2 )}
\end{equation}
where the real numbers $\mu_1$ and $\mu_2$ replace as thermodynamic parameters
the densities $\rho_1$ and $\rho_2$. The corresponding specific free energy,
the pressure, is the limit
\begin{equation}
  \label{pressure} p ( \mu_1, \mu_2 ) = \lim_{\Lambda \uparrow \mathbbm{Z}^d}
  \frac{1}{| \Lambda |} \ln \Xi_{bc} ( \Lambda ; \mu_1, \mu_2 )
\end{equation}
Limits (\ref{freeenergy}) and (\ref{pressure}), which define the above free
energies, exist. They are convex functions of their parameters and are related
by the Legendre transformations
\begin{eqnarray}
  p ( \mu_1, \mu_2 ) & = & \sup_{\rho_1, \rho_2} [ \mu_1 \rho_1 + \mu_2 \rho_2
  - \beta f ( \rho_1, \rho_2 ) ] \\
  \beta f ( \rho_1, \rho_2 ) & = & \sup_{\mu_1, \mu_2} [ \mu_1 \rho_1 + \mu_2
  \rho_2 - p ( \mu_1, \mu_2 ) ] 
\end{eqnarray}

Finally we introduce the finite volume Gibbs measures (a specification)
associated with the second ensemble:
\begin{equation}
  \label{eq:gm} \mathbb{P}^{bc}_{\Lambda} ( \mathbf{s}_{\Lambda} ) =
  \frac{e^{- \beta \tilde{H}_{\Lambda} ( \mathbf{s}_{\Lambda} )} \chi^{bc} (
  \mathbf{s}_{\Lambda} )}{\Xi_{bc} ( \Lambda ; \mu_1, \mu_2 )}
\end{equation}
where $\tilde{H}_{\Lambda} ( \mathbf{s}_{\Lambda} ) = H_{\Lambda} (
\mathbf{s}_{\Lambda} ) - \frac{\mu_1}{\beta} \sum_{x \in \Lambda} \delta (
s_x, 1 ) - \frac{\mu_2}{\beta} \sum_{x \in \Lambda} \delta ( s_x, 2 )$. They
determine by the Dobrushin--Landford--Ruelle equations the set of Gibbs states
$\mathcal{G_{\beta} ( \tilde{H} )}$ on $\mathbb{Z}^d$ corresponding to the
Hamiltonian $\tilde{H}$ at inverse temperature $\beta$ (see e.g. {\cite{R2}}).
A Gibbs state $\mathbb{P} \in \mathcal{G_{\beta} ( \tilde{H} )}$ which equal
the limit $\lim_{\Lambda \uparrow \mathbb{Z}^d} \mathbb{P}^{bc}_{\Lambda} (
\cdot )$, is called  Gibbs state with boundary condition $b.c$.

\vspace{.2cm}
\noindent
\textbf{\textit{Ground states and low temperature analysis}}

In the zero temperature limit the Gibbs state with empty boundary condition is
concentrated on the configuration with empty sites:
\begin{equation}
  \lim_{\beta \rightarrow \infty} \mathbb{P}^{bc}_{\Lambda} (
  \text{emp}_{\Lambda} ) = 1
\end{equation}
where $\text{emp}_{\Lambda}$ is the configuration where all the sites of
$\Lambda$ are empty, and this limit vanishes for any other configuration.
Gibbs states at $\beta = \infty$ are called ground states.

Let, $R^{_{\text{mixt}}}_{\Lambda} ( c ) = \left\{ s \in \Omega^{\Lambda} :
\forall x \in \Lambda, s_x \not{=} 0 \right\}$, be the \textit{restricted
ensemble} of configurations in $\Lambda$ with non empty sites, and
$R^{_{\text{mixt}}}_{\Lambda} ( c )$, $0 \leq c \leq 1$ the subset of
configurations of $R^{_{\text{mixt}}}_{\Lambda}$ with exactly $[ c| \Lambda |
] = N$ sites occupied by a particle of the specie $1$ (and $| \Lambda | - [ c|
\Lambda |$ sites occupied by a particle of the specie $2$).

With the mixture boundary conditions one gets by by Stirling's approximation
and the principle of maximal term that
\begin{equation}
  \label{eq:gsmixt1} 
  \lim_{\beta \rightarrow \infty}
  \mathbb{P}^{\text{mixt}}_{\Lambda} ( R^{_{\text{mixt}}}_{\Lambda} ( c ) )
  \build{\longrightarrow }_{\Lambda \rightarrow \mathbbm{Z}^d }^{}
  1
\end{equation}
for $c = \frac{e^{\mu_1}}{e^{\mu_1} + e^{\mu_2}}$. This means that the ground
state with $\text{mixt}$ boundary conditions is concentrated on the restricted
ensemble $R^{_{\text{mixt}}} ( c )$ of configurations of non empty sites with
concentration $c$ of particles $1$ and concentration $1 - c$ of particles $2$.

With free boundary conditions, one has for $c = \frac{e^{\mu_1}}{e^{\mu_1} +
e^{\mu_2}}$
\[ 
\lim_{\beta \rightarrow \infty} \mathbb{P}^{fr}_{\Lambda} (
   \text{emp}_{\Lambda} ) = 1 / 2, \lim_{\beta \rightarrow \infty}
   \mathbb{P}^{fr}_{\Lambda} ( R^{_{\text{mixt}}}_{\Lambda} ( c ) )
  \build{\longrightarrow }_{\Lambda \rightarrow \mathbbm{Z}^d }^{}
    1 / 2
     \]
Thus, with the above considerations, we get that for $e^{\mu_1} + e^{\mu_2} =
1$, the configuration with empty sites coexists with the restricted ensemble
$R^{_{\text{mixt}}} ( c )$.

This analysis can be extend to the Gibbs states at low temperatures by using
Pirogov-Sinai theory {\cite{S}}. Actually, this theory allows to show that the
low temperature phase diagram of the model is a small perturbation of the
diagram of ground states the coexistence line given by the equation
\begin{eqnarray*}
  \ln ( { e^{{ \mu_1^{\ast}}} + 
  e^{{
  { { \mu_2^{\ast}}}}}} ) & = & e^{{
  { { \mu_1^{\ast}}}} - 2 d { \beta
  J_1}} + e^{{{ \mu_2^{\ast}}} - 2 d {
  \beta J_2}}\\
  &  & \begin{array}{ll}
    - & \frac{( e^{{ \mu_1^{\ast} - \beta J_1}}
    + e^{{ \mu_2^{\ast} - \beta J_2}} )^{2 d}}{(
    e^{{ { \mu_1^{\ast}}}} + e^{{
    { \mu_2^{\ast}}}} ) ^{2  d + 1}}
  \end{array} \begin{array}{ll}
    + & O \left( e^{- ( 2 d + 1 ) { \beta J} {
    }}  \right)
  \end{array}
\end{eqnarray*}
where ${ J = \min \{ J_{1}, J_2 \}}$.

In particular, introducing the infinite volume expectation $\langle
\hspace{0.25em} \cdot \hspace{0.25em} \rangle^{\text{bc}} ( \mu_1, \mu_2 )$
associated to the Gibbs measure (\ref{eq:gm}), $\langle \hspace{0.25em} \cdot
\hspace{0.25em} \rangle^{\text{bc}} ( \mu_1, \mu_2 ) = \lim_{\Lambda \uparrow
\mathbbm{Z}^d} \sum_{\mathbf{s}_{\Lambda} \in \Omega^{\Lambda}} \cdot
\hspace{0.25em} \mathbbm{P}_{\Lambda}^{\text{bc}} ( \mathbf{s}_{\Lambda} )$,
we have for any $t \geq 0$:
\begin{eqnarray}
  \langle \delta ( s_x, 1 ) + \delta ( s_x, 2 ) \rangle^{\text{mixt}} (
  \mu_1^{\ast} + t, \mu_2^{\ast} + t ) & \geq & 1 - O \left( e^{- 2 d \beta J}
  \right) \nonumber\\
  \langle \delta ( s_x, 1 ) + \delta ( s_x, 2 ) \rangle^{\text{emp}} (
  \mu_1^{\ast} - t, \mu_2^{\ast} - t ) & \leq & O \left( e^{- 2 d \beta J}
  \right) \nonumber
\end{eqnarray}
showing that the model exhibits at low temperature a first order phase
transition at the coexistence line where the pressure is discontinuous.

\vspace{.2cm}
\noindent
\textbf{\textit{Surface tensions}}

To introduce the surface tension between the mixture and the vapor, we
consider the parallelepipedic box:
\[ V = V_{L, M} = \left\{ ( x_1, .., x_d ) \in \mathbb{Z}^d : |x_i | \leq L, i
   = 1, ..., d - 1 ; - M \leq x_d \leq M - 1 \right\} \]
and let $\partial_+ V$ (respectively $\partial_- V$) be the set of sites of
$\partial V$ with $x_d \geq 0$ (respectively $x_d < 0$). The boundary
condition, $\chi^{\text{mixt}, \text{emp}} ( \mathbf{s}_V ) = \prod_{x \in
\partial_- V} ( 1 - \delta ( s_x, 0 ) ) \prod_{x \in \partial_+ V} \delta (
s_x, 0 )$, enforces the existence of an interface between the mixture and the
vapor and the interfacial tension between the mixture and the vapor is defined
by the limit
\[ \label{eq:surtenmel} \tau_{( 1, 2 ) |0} = - \frac{1}{\beta} \lim_{L
   \rightarrow \infty, M \rightarrow \infty} \tmop{li}_{} \frac{1}{( 2 L + 1
   )^{d - 1}} \ln \frac{\Xi_{\text{mixt}, \text{emp}} ( V ; \mu^{\ast}_1,
   \mu^{\ast}_2 )}{( \Xi_{\text{mixt}} ( V ; \mu^{\ast}_1, \mu^{\ast}_2 )
   \Xi_{\text{emp}} ( V ; \mu^{\ast}_1, \mu^{\ast}_2 ) )^{1 / 2}} \]
As mentioned previously, whenever either the particles $1$ or the particles
$2$ are not allowed the system reduces to the usual Ising model in its lattice
gas version. Thus, to define the surface tensions between each species of the
mixture and the vapor, we introduce the configurations $n_V \in \left\{ 0, 1
\right\}^V$ of the lattice gas and the following partition functions
\begin{eqnarray*}
  Q_{\alpha} ( V ) & = & \sum_{n_V \in \left\{ 0, 1 \right\}^V} e^{\beta
  J_{\alpha} \sum_{\langle x, y \rangle \subset V} [ n_x ( 1 - n_y ) + ( 1 -
  n_x ) n_y ]} \prod_{x \in \partial V} n_x\\
  Q_{\alpha, 0} ( V ) & = & \sum_{n_{\Lambda} \in \left\{ 0, 1 \right\}^V}
  e^{\beta J_{\alpha} \sum_{\langle x, y \rangle \subset V} [ n_x ( 1 - n_y )
  + ( 1 - n_x ) n_y ]} \prod_{x \in \partial_- V} ( 1 - n_x ) \prod_{x \in
  \partial_+ V} n_x
\end{eqnarray*}
for $\alpha = 1$ and $\alpha = 2$. The interfacial tension between the species
$\alpha = 1, 2$, and the vapor is the limit ({\cite{GMM,BLP2}})
\[ \tau_{\alpha, 0} = \lim_{L \rightarrow \infty} \lim_{M \rightarrow \infty}
   \frac{F_{\alpha} ( V )}{( 2 L + 1 )^{d - 1}} \]
where $F_{\alpha} ( V ) = - \frac{1}{\beta} \ln \frac{Q_{\alpha, 0} ( V
)}{Q_{\alpha} ( V )}$. It is well known that the ratio $Q_{\alpha, 0} ( V ) /
Q_{\alpha} ( V )$ can be expressed as a sum over interfaces which in this case
are connected set of bonds or plaquettes of the dual lattice {\cite{G,D1}}.
Extracting the energy of the flat interface, the system can be written as a
gas of excitations leading to $F_{\alpha} ( V ) = J_{\alpha} ( 2 L + 1 )^{d -
1} + F_{\alpha}^{\text{ex}} ( V )$. In two dimensions,
$F_{\alpha}^{\text{ex}}$ is the free energy of the gas of jumps of the
Gallavotti's line {\cite{G}}. In three dimensions, $F_{\alpha}^{\text{ex}}$ is
the free energy of the gas of walls of the Dobrushin's interface {\cite{D1}}.
In both cases these free energies can be analyzed by cluster expansion
techniques at low temperatures. Namely, the specific free energies
$\mathcal{F}_{\alpha} = \lim_{L \rightarrow \infty} F_{\alpha}^{\text{ext}} (
V ) / ( 2 L + 1 )^{d - 1}$ exist and are given by convergent expansions in
term of the activities $e^{- \beta J_{\alpha}}$, giving
\begin{equation}
  \label{eq:tjf} \tau_{\alpha, 0} = J_{\alpha} + \mathcal{F}_{\alpha}
\end{equation}
In addition
\begin{eqnarray}
  - \beta \mathcal{F}_{\alpha} & = & 2 e^{- \beta J_{\alpha}} + O ( e^{- 2
  \beta J_{\alpha}} ) \quad \text{for} \quad d = 2  \label{eq:stim2}\\
  - \beta \mathcal{F}_{\alpha} & = & 2 e^{- 4 \beta J_{\alpha}} + O ( e^{- 6
  \beta J_{\alpha}} ) \quad \text{for} \quad d = 3  \label{eq:stim3}
\end{eqnarray}
Furthermore, in two dimensions the surface tension defined above is known to
coincide with the one computed by Onsager {\cite{GMM}}. We thus have an exact
expression for $\tau_{\alpha, 0}$, and for $\mathcal{F}_{\alpha}$, namely,
$\beta \mathcal{F}_{\alpha} = \ln \tanh ( \beta J_{\alpha} / 2 )$for $\beta
J_{\alpha}$ larger than the critical value $\ln ( 1 + \sqrt{2} )$.

\vspace{.2cm}
\noindent
\textbf{\textit{Main result}}

The above surface tensions $\tau_{( 1, 2 ) |0}$, $\tau_{1, 0}$ and $\tau_{2,
0}$, are proved to satisfy, whenever $\beta$ is large enough, the equation
{\cite{DMR}}:
\begin{equation}
  e^{- \beta ( \tau_{( 1, 2 ) |0} - \mathcal{F} )} = c^{\ast}_1 e^{- \beta (
  \tau_{1|0} - \mathcal{F}_1 )} + c^{\ast}_2 e^{- \beta ( \tau_{2|0} -
  \mathcal{F}_2 )_{}}
\end{equation}
The quantity $\mathcal{F}$ is the specific free energy (which can be expressed
as a convergent series at low temperatures) of a gas of some geometrical
objects called aggregates. In dimension $d = 2$, those aggregates are the
natural generalizations to our model of the jumps of Gallavotti's line and the
leading term of the series giving this free energy $\mathcal{F}$ is
\[ - \frac{2}{\beta} \frac{c^{\ast}_1 e^{- 2 \beta J_1} + c^{\ast}_2 e^{- 2
   \beta J_2}}{c^{\ast}_1 e^{- \beta J_1} + c^{\ast}_2 e^{- \beta J_2}} \]
In dimension $d = 3$, they are the natural generalizations of the walls of the
Dobrushin's interface and then the leading term of the series is
\[ - \frac{1}{\beta} \frac{c^{\ast}_1 e^{- 5 \beta J_1} + c^{\ast}_2 e^{- 5
   \beta J_2}}{c^{\ast}_1 e^{- \beta J_1} + c^{\ast}_2 e^{- \beta J_2}} -
   \frac{1}{\beta} \frac{( c^{\ast}_1 e^{- 2 \beta J_1} + c^{\ast}_2 e^{- 2
   \beta J_2} )^4}{( c^{\ast}_1 e^{- \beta J_1} + c^{\ast}_2 e^{- \beta J_2}
   )^4} \]
The coefficients $c^{\ast}_1$ and $c^{\ast}_2$ are related to the
concentrations $c_1$ and $c_2$ of the particles $1$ and the particles $2$
through the equation
\[ c^{\ast}_i = e^{\mu_i^{\ast} - p ( \mu_1^{\ast}, \mu_2^{\ast} )}, i = 1, 2
\]
This equation gives at low temperatures:
\begin{eqnarray}
  c^{\ast}_i & = & c_i \left[ 1 - \left( c_1 e^{- \beta J_1} + c_2 e^{- \beta
  J_2} \right)^{2 d} - 2 dc_i e^{- \beta J_i} \left( c_1 e^{- \beta J_1} + c_2
  e^{- \beta J_2} \right)^{2 d - 1} \right. \nonumber\\
  &  & \left. - 2 ( d + 1 ) c_i \left( c_1 e^{- \beta J_1} + c_2 e^{- \beta
  J_2} \right)^{2 d} + O \left( e^{- ( 2 d + 1 ) \beta \min \{ J_1, J_2 \}}
  \right) \right] 
\end{eqnarray}


\begin{thebibliography}{10}
  \bibitem[1]{DR}J. De Coninck and J. Ruiz. {\newblock}Interfacial tensions
  for binary mixtures versus Solid--On--Solid models. {\newblock}
  \textit{Preprint}, 2003.
  
  \bibitem[2]{DMR}J. De Coninck, S. Miracle--Sole, and J. Ruiz. {\newblock}On
  the statistical mechanics and surface tensions of binary mixtures.
  {\newblock} \textit{Preprint CPT--Marseille--2004/P.015, mparc 04--206,
  submitted to Journal of Statistical Physics}.
  
  \bibitem[3]{Gu}E. A. Guggenheim. {\newblock} \textit{Trans. Faraday Soc.},
  \textbf{41}:150, 1945.
  
  \bibitem[4]{DP}R. Defay and I. Prigogine. {\newblock} \textit{Trans. Faraday
  Soc.}, \textbf{46}:199, 1950.
  
  \bibitem[5]{Eb}J. G. Eberhart. {\newblock} \textit{J. Phys. Chem.},
  \textbf{70}:1183, 1966.
  
  \bibitem[6]{Sz}B. von Szyszkowsky. {\newblock} \textit{J. Phys. Chem.},
  \textbf{64}:385, 1908.
  
  \bibitem[7]{Sz2}H. P. Meissner and A. S. Michaels. {\newblock} \textit{Ind.
  Eng. Chem.}, \textbf{42}:2782, 1949.
  
  \bibitem[8]{Ad}A. W. Adamson. {\newblock} \textit{Physical Chemistry of
  Surfaces}. {\newblock}John Wiley and Sons, New--York, 1997.
  
  \bibitem[9]{DPBE}R. Defay, I. Prigogine, A. Bellmans, and D. H Everett.
  {\newblock} \textit{Surface tensions and absorption}. {\newblock}Longmans,
  Green and Co., London, 1966.
  
  \bibitem[10]{WW}J. C. Wheeler and B. Widom. {\newblock} \textit{J. Chem.
  Phys.}, \textbf{52}:5334, 1970.
  
  \bibitem[11]{LG}J. L. Lebowitz and G. Gallavotti. {\newblock}Phases
  transitions in binary lattice gases. {\newblock} \textit{J. Math. Phys.},
  \textbf{12}:1129, 1971.
  
  \bibitem[12]{R1}D. Ruelle. {\newblock} \textit{Statistical Mechanics:
  Rigorous Results}. {\newblock}Benjamin, New--York, Amsterdam, 1969.
  
  \bibitem[13]{R2}D. Ruelle. {\newblock} \textit{Thermodynamic Formalism}.
  {\newblock}Addison Wesley, Reading, 1978.
  
  \bibitem[14]{S}Ya. G. Sinai. {\newblock} \textit{Theory of Phase
  Transitions: Rigorous Results}. {\newblock}Pergamon Press, London, 1982.
  
  \bibitem[15]{GMM}A. Martin Löf, G. Gallavotti, and S. Miracle--Sole.
  {\newblock}Some problem connected with the coexistence of phase in the Ising
  model. {\newblock}In \textit{Statistical mechanics and mathematical
  problems, Lecture Notes in Physics}, volume 20, page 162. Springer, Berlin,
  1973.
  
  \bibitem[16]{BLP2}J. Bricmont, J. L. Lebowitz, and C.--E. Pfister.
  {\newblock}On the surface tensions of lattice systems. {\newblock}
  \textit{Ann. N. Y. Acad. Sci.}, \textbf{337}:214, 1980.
  
  \bibitem[17]{G}G. Gallavotti. {\newblock}Phase separation line in the
  two--dimensional Ising model. {\newblock} \textit{Commun. Math. Phys.},
  \textbf{27}:103, 1972.
  
  \bibitem[18]{D1}R. L. Dobrushin. {\newblock}Gibbs states describing the
  coexistence of phases for a three-dimensional Ising model. {\newblock}
  \textit{Theory Prob. Appl.}, \textbf{17}:582, 1972.
\end{thebibliography}
\end{document}